\documentclass[aps,prb,twocolumn,superscriptaddress,amsmath,amssymb,floatfix,longbibliography]{revtex4-1}
\usepackage{bm}
\usepackage{xcolor}
\usepackage{graphicx}
\usepackage{lineno}
\usepackage{comment}
\usepackage[caption=false]{subfig} 

\def\efield{\boldsymbol{\mathcal{E}}} 
\newcommand{\PP}{\textbf{P}}
\newcommand{\kk}{\textbf{k}}

\begin{document}
\title{How strong is the Second Harmonic Generation in single-layer monochalcogenides?\\
A response from first-principles real-time simulations}

\newcommand{\cinam}{CNRS/Aix-Marseille Universit\'e, Centre Interdisciplinaire de Nanoscience de Marseille UMR 7325 Campus de Luminy, 13288 Marseille Cedex 9, France}
\newcommand{\qub}{School of Mathematics and Physics, Queen's University Belfast, Belfast BT7 1NN, Northern Ireland, UK}
\newcommand{\etsf}{European Theoretical Spectroscopy Facilities (ETSF)}
\newcommand{\torvergata}{University of Rome Tor Vergata, Rome, Italy}
\newcommand{\piim}{Universit\'e Aix-Marseille, Laboratoire de Physique des Interactions Ioniques et Moléculaires (PIIM), UMR CNRS 7345, F-13397 Marseille, France}

\author{Claudio Attaccalite}
\affiliation{\cinam}
\affiliation{\etsf}

\author{Maurizia Palummo}
\affiliation{\torvergata}
\affiliation{\etsf}
\author{Elena Cannuccia}
\affiliation{\torvergata}
\affiliation{\piim}
\author{Myrta Gr\"uning}
\affiliation{\qub}
\affiliation{\etsf}

\begin{abstract}

Second Harmonic Generation (SHG) of single-layer monochalcogenides, such as  GaSe and InSe, has been recently reported [2D Mater. 5 (2018) 025019; J. Am. Chem. Soc. 2015, 137, 7994−7997] to be extremely strong with respect to bulk and multilayer forms. To clarify the origin of this strong SHG signal, we perform first-principles real-time simulations of linear and non-linear optical properties of these two-dimensional semiconducting materials. The simulations, based on ab-initio many-body theory, accurately treat the electron-hole correlation and capture excitonic effects that are deemed important to correctly predict the optical properties of such systems. We find indeed that, as observed for other 2D systems, the SHG intensity is redistributed at excitonic resonances. The obtained theoretical SHG intensity is an order of magnitude smaller than that reported at the experimental level. This result is in substantial agreement with previously published simulations which neglected the electron-hole correlation, demonstrating that many-body interactions are not at the origin of the strong SHG measured. We then show that the experimental data can be reconciled with the theoretical prediction when a single layer model, rather than a bulk one, is used to extract the SHG coefficient from the experimental data.
\end{abstract}           

\maketitle

\section{Introduction}
Nowadays there is a considerable interest in the excited state properties of 2D materials which can be distinctly different from that of their bulk counterpart.\cite{doi:10.1002/adma.201705963} 
Strongly bound excitons often characterize the optical spectra of two-dimensional (2D) semiconducting crystals and in several cases,  intense nonlinear optical spectra have been observed ~\cite{Wang2015,saynatjoki2017ultra} 
such as  strong second-harmonic generation~\cite{PhysRevB.87.201401,PhysRevB.87.161403,Janisch2014,2053-1583-2-4-045015,zhou2015strong} (SHG)---typically up to one order of magnitude larger than in conventional nonlinear crystals. For this reason these materials are of potential technological interest as frequency converters in nanophotonic circuits.~\cite{Majumdar2015}
Furthermore, the SHG is an excellent spectroscopic tool for the imaging and characterization of 2D materials~\cite{Li2013,saynatjoki2017ultra,PhysRevB.87.201401,PhysRevB.87.161403,zhou2015strong,Hsu2014,Yin2014} and understand the origin of the signal from theoretical point of view is of fundamental importance for the interpretation.  

Recently, extremely large values for the SHG of single-layer monochalcogenides have been reported: for InSe~\cite{zhou2018inse} a value of 6.39~nm/V at 1.55~eV and for GaSe~\cite{zhou2015strong} a value of 2.4~nm/V at 1.02~eV. To put these values into perspective, the SHG of single-layer MoS$_2$ is of the order of 0.1--0.4 nm/V~\cite{PhysRevB.87.201401,PhysRevB.90.121409} at resonance, that of bi- and tri- and multilayer~\cite{jie2015layer,PhysRevB.94.125302,karvonen2015investigation} GaSe ranges from 9--90 pm/V, and that of bulk GaSe is about 60~pm/V. A similar large SHG of 10~nm/V was reported for WS$_2$ and WSe$_2$ monolayers.~\cite{Janisch2014,2053-1583-2-4-045015} However, the extremely large SHG coefficients for single-layer monochalcogenides are not confirmed by first-principle calculations\cite{lin2018indium,hu2017layer} which give consistently values smaller by one order of magnitude.

The discrepancy between the theoretical and experimental results may be due to many-body effects that were neglected in the computational studies of Refs.~\onlinecite{lin2018indium,hu2017layer}. In fact, electron-electron interaction and excitonic effects are known to play a crucial role in the description of optical properties, especially in the case  of low-dimensional materials. Indeed the incomplete screening of the electron-electron interaction and the quantum confinement may further enhance these effects with respect to bulk. Specifically for single-layer monochalcogenides, the formation of strong bound excitons have been recently shown in Ref.~\onlinecite{antonius2018orbital}. Further, their peculiar optical properties have been attributed to saddle-points excitations\cite{luo2018saddle} that originate from the Mexican hat-like band structure near the $\Gamma$ point.\cite{rybkovskiy2014transition,chen2018large} 
\begin{figure}[h]
\centering
    \includegraphics[width=0.45\textwidth]{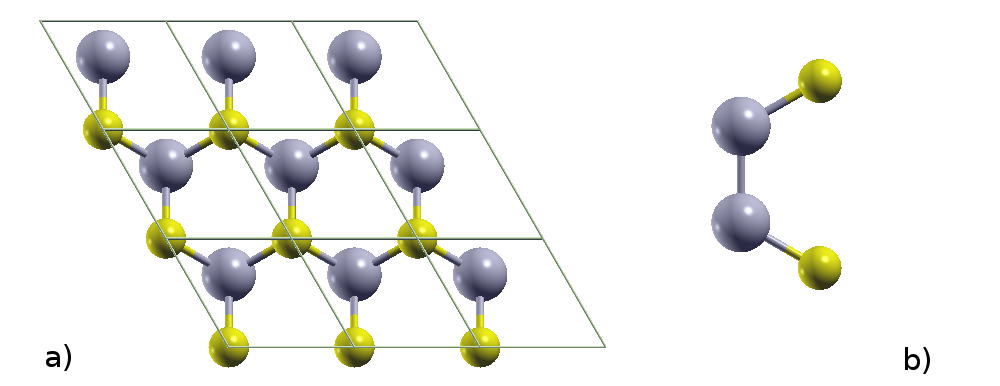}
     \caption{\footnotesize{[Color online] Ball-and-stick representation of the atomic structure of the GaSe (InSe) monolayer: top view (a) and lateral view (b). Gallium (Indium) atoms are in gray, Selenium in yellow.\label{atoms}}}
\end{figure}

This work addresses the question of whether the extremely large SHG of single-layer monochalcogenides is due to strongly bound excitons. 
To this end, by using a first-principle real-time approach based on many-body theory (Sec.~\ref{sc:theo}), we calculate the electronic structure, dielectric function (Sec.~\ref{sc:res1}) and the SHG (Sec.~\ref{sc:res2}) of InSe and GaSe monolayers and, in order to single out many-body effects, we compare with the results at the independent-particle level. Previous studies on h-BN, MoS$_2$  and other semiconducting 2D materials found a significant enhancement at resonance of the SHG signal due to excitonic effects---up to 2--4 times depending on the system.~\cite{Gruning2014, Attaccalite2015} For single-layer monochalcogenides, we find that intensity is only slightly redistributed by many-body effects respect to  the independent-particle approximation. We conclude (Sec.~\ref{sc:conc}) that many-body effects do not account for the difference between theoretical predictions\cite{lin2018indium,hu2017layer} and experimental estimates.\cite{zhou2018inse,zhou2015strong} We turn then the attention to the experimental results and argue that the extremely large SHG coefficient is an artifact of the model assumed to extract the estimate from the experimental data.

\section{Theoretical framework and computational details}\label{sc:theo}
The structural optimization and electronic structure have been calculated using density-functional theory (DFT). The crystal structure of the GaSe and InSe monolayers (Fig.~\ref{atoms}) has been obtained from the corresponding bulk structure using the same in-plane lattice parameters ($a=0.3743$~nm and $a=0.4$~nm respectively for the GaSe~\cite{schwarz2007effect} and the InSe monolayer~\cite{de1982large}). Since experiments on GaSe and InSe were performed on different substrates, the lattice constant of these materials could depend on the substrate interaction. Therefore we decided to keep fixed the lattice constants to the bulk values and optimizing only the atomic positions that are not fixed by symmetries, namely the $z$ coordinates of the atoms. In order to simulate an isolated monolayer, we used a 35 a.u. supercell in the $z$-direction. 

DFT calculations have been performed with the Quantum-Espresso code\cite{pwscf} using the Perdew-Burke-Ernzerhof\cite{PhysRevLett.77.3865} (PBE) functional and the scalar-relativistic optimized norm-conserving Vanderbilt\cite{PhysRevB.88.085117} pseudopotentials from the PseudoDojo repository (v0.4).\cite{PseudoDojo} The valence configuration for each pseudo-atoms are: $3d^{10}4s^2~4p^1$ for the Ga, $4d^{10}5s^2~5p^1$ for the In and $4s^24p^4$ for Se. We used a shifted $18 \times 18 \times 1$ $\kk$-point sampling for the ground-state, a plane-wave cutoff of $90$~Ry for the structural optimization and $70$~Ry for the band-structure calculations.

Calculations of the quasi-particle energies and optical susceptibilities have been performed using the Yambo code.\cite{yambo} The quasiparticle corrections to the fundamental band gap have been calculated from the standard $G_0W_0$ approximation\cite{lucia} with the Godby-Needs plasmon-pole model\cite{godby1989PPA} and applied as a rigid shift to all the bands. We used a $24 \times 24 \times 1$ $\kk$-point sampling, a 5~Hartree cutoff for the dielectric function, and 200 total bands for Green's function expansion. We calculate the gap correction at $\Gamma$ point and then shifted all conduction bands by this amount. In this way the quasiparticle band structure has the same band ordering and band-width of the DFT one, with the only difference that the gap has been corrected within the $G_0W_0$ approximation. Convergence with the number of conduction bands was accelerated by means of the Bruneval-Gonze terminator.\cite{bruneval2008accurate} Optical absorption spectra have been obtained by the solution of the Bethe-Salpeter equation\cite{strinati} using a basis of electron-hole pairs for which we considered 6 valence and 7 conduction states for both GaSe and InSe monolayers. 
We verified that increasing the number of bands in the calculations does not change the spectra in the energy range considered here.

\begin{figure}[ht]
\centering
\includegraphics[width=0.45\textwidth]{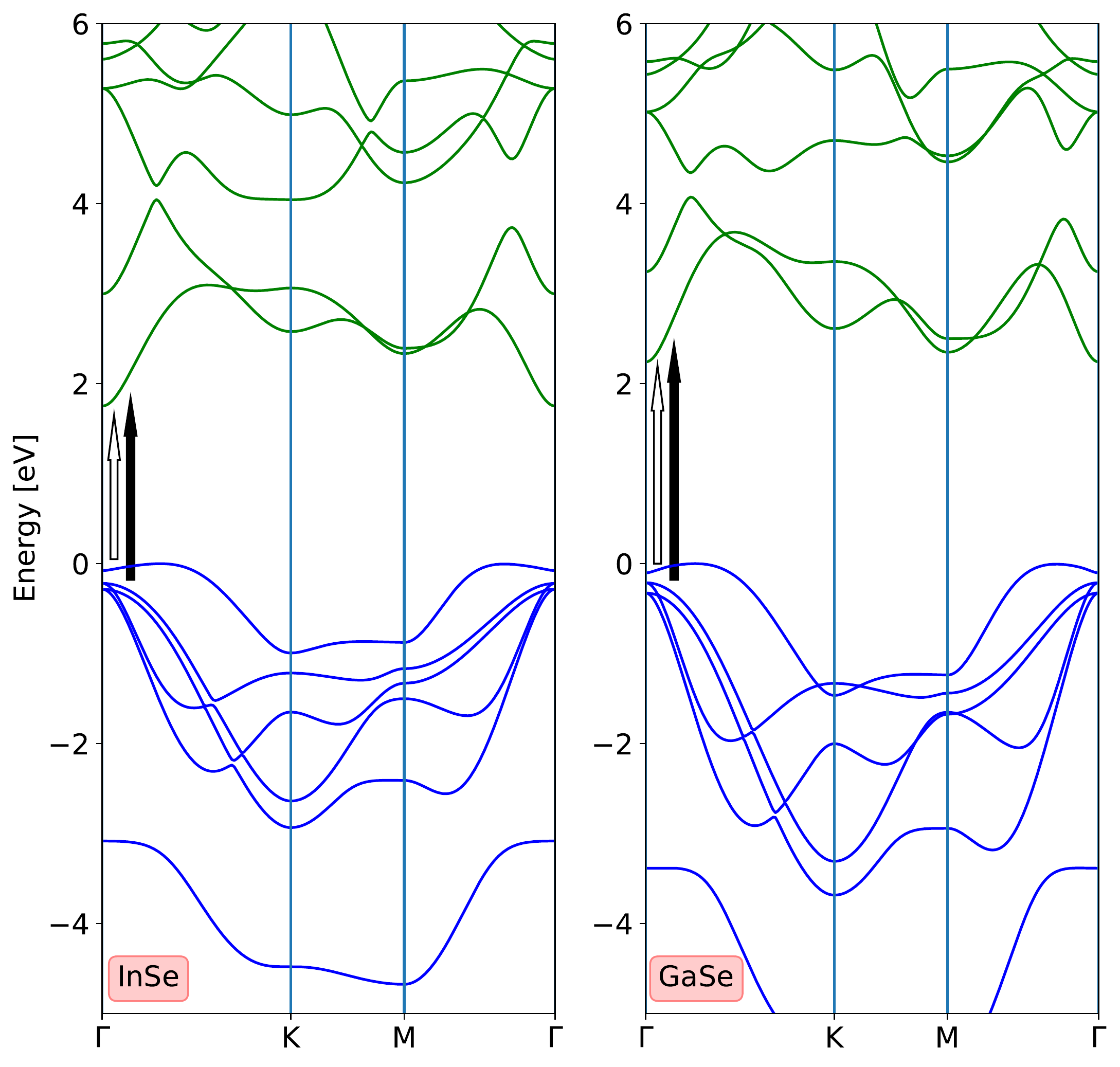}
     \caption{\footnotesize{[Color online] On the left (right) the electronic band structure of InSe (GaSe) monolayer at the DFT level. Valence bands are in blue, conduction bands in green. For in-plane polarized light, the white arrow indicates the dipole-forbidden transition between the top-valence and bottom conduction bands,  while the black arrow is the lowest dipole-allowed transition from the second-top valence and the bottom conduction band. \label{bandsGaSe}}}
\end{figure}
Non-linear susceptibilities have been obtained from the real-time evolution of Bloch-electrons in a uniform time-dependent electric field following the approach proposed in Ref.~\onlinecite{Attaccalite2013, Gruning2014}.
The effective Hamiltonian for the Bloch-electrons, derived from many-body theory, contains both the electron-hole attraction (through a screened-Coulomb term) and exchange (through a Hartree term) needed to describe excitons and local-field depolarization effects. The whole framework is based on DFT and quasiparticle corrections are included at the level of $G_0W_0$.
In the linear-response limit, the approach is equivalent to the solution of the BSE on top of the $G_0W_0$ electronic structure~\cite{attaccalite2011real} and we therefore refer to this level of theory as $G_0W_0+$BSE for both linear and nonlinear susceptibilities. Within this framework, we can exclude specific terms from the effective Hamiltonian to investigate how they affect the SHG. In particular, we consider the independent-particle approximation (IPA)---no electron-hole correlation; the random-phase approximation (RPA)---only electron-hole exchange. Further, we perform the simulations on top of either the DFT or the $G_0W_0$ electronic structure. For the screened-Coulomb and Hartree terms, we use the same parameters as in the BSE calculations.   

After the integration of the equation-of-motion, the nonlinear susceptibilities $\chi^{(n)}$ are extracted using Fourier techniques from the macroscopic polarization $\PP$---calculated as a dynamic Berry-phase~\cite{Souza2004}---and its expansion in power of the total electric field $\efield$:
\begin{equation}
\PP_i = \chi^{(1)}_{ij} \efield_j + \chi^{(2)}_{ijk}  \efield_j \efield_k +  \chi^{(3)}_{ijkl} \efield_j \efield_k \efield_l + O(\efield^4)  \; ,
\end{equation}
where the $i,j,k,l$ subscripts refer to Cartesian components of the field and polarization.

Specifically, to obtain a single SHG spectrum we perform a series of simulations of about 75~fs for a monochromatic electric field with frequency ranging from 0.1~eV to 7.0~eV. We integrate the equation-of-motion using the Crack-Nicholson method~\cite{crank1947practical} with a time-step of 0.01~fs. 
Since the turning-on of the electric-field excites all eigenfrequencies of the system, we introduce a phenomenological decoherence non-Hermitian operator corresponding to a Lorentzian broadening of the spectrum of 0.2 eV.~\footnote{It can be shown in Ref.\onlinecite{wismer2018gauge} that for a simple decoehrence operator, as the one used here, there is an equivalence between the real-time dynamics generated by a non-Hermitian Hamiltonian and the dephasing usually employed in the density matrix formalisms. However, the Hamiltonian formalism is more convenient in our case because the polarization can be calculated directly from the bloch-electrons evolution. The intensity of SHG depends on this phenomenological parameter and we chose a value compatible with experimental broadening.Another possibility would be to calculate the dephasing by including the coupling with the atomic motion explicitely in our simulations as done in Ref.~\onlinecite{monserrat2018electron}.} The contribution of the eigenfrequencies to the signal decays exponentially in time with the decay constant being the inverse of the dephasing time. After five times the dephasing time (about 33~fs) the eigenfrequency signal is negligible with respect to the SHG signal and we then perform the Fourier analysis as detailed in Ref.~\onlinecite{Attaccalite2013}.

The static limit ($\omega=0$), corresponding to the outmost left point of Figs.~\ref{eps},\ref{epsInSe} has been obtained by extrapolation of the values at small frequencies of the $\chi^{(2)}(\omega)$.

The external electric field is polarized along the $y$ direction and the polarization is recorded in the same direction, obtaining the $\chi^{(2)}_{yyy} = -\chi^{(2)}_{xxy} = -\chi^{(2)}_{yyx} = -\chi^{(2)}_{xyx} = -\chi^{(2)}_{aab} $  that is the only nonzero component of the second-order susceptibility tensor for the hexagonal $D_{3h}$ crystal class,~\cite{Boyd2008} being $a$ and $b$ the in-plane crystal axes. 
To obtain a SHG signal independent of the dimension of the supercell, we rescaled the calculated $\chi^{(2)}(\omega)$ by an effective thickness of $0.796$~nm for the GaSe and $0.832$~nm for the InSe---corresponding to half of $c$ lattice parameter of the bulk structures.~\cite{Boyd2008}

%
%
\section{Results}\label{sc:res}
\subsection{Electronic structure and optical properties}\label{sc:res1}
In Fig.~\ref{bandsGaSe}, we report the electronic band structure of both monochalcogenide monolayers and the position of the lowest excitons obtained from the solution of the BSE. Because of the mirror-plane symmetry ($z \rightarrow -z$), electronic states near the edges are either even or odd with respect to this symmetry operation.\cite{zolyomi2014electrons} As a result, the lowest-energy electron-‐hole transition---depicted by a white arrow---is optically inactive for in-plane polarized light and only slightly coupled to $z$-polarized light.\cite{antonius2018orbital,bandurin2017high} In the same figure, we also indicate the lowest optical active transition with a black arrow.

The band structure is in good agreement with previous results obtained by Hu et al.\cite{hu2017layer}, while we found small differences respect to Antonius et al.\cite{antonius2018orbital} who reported a larger energy difference between the two top valence bands at the vicinity of $\Gamma$. 

\begin{figure*}[ht]
\centering
\includegraphics[width=0.99\textwidth]{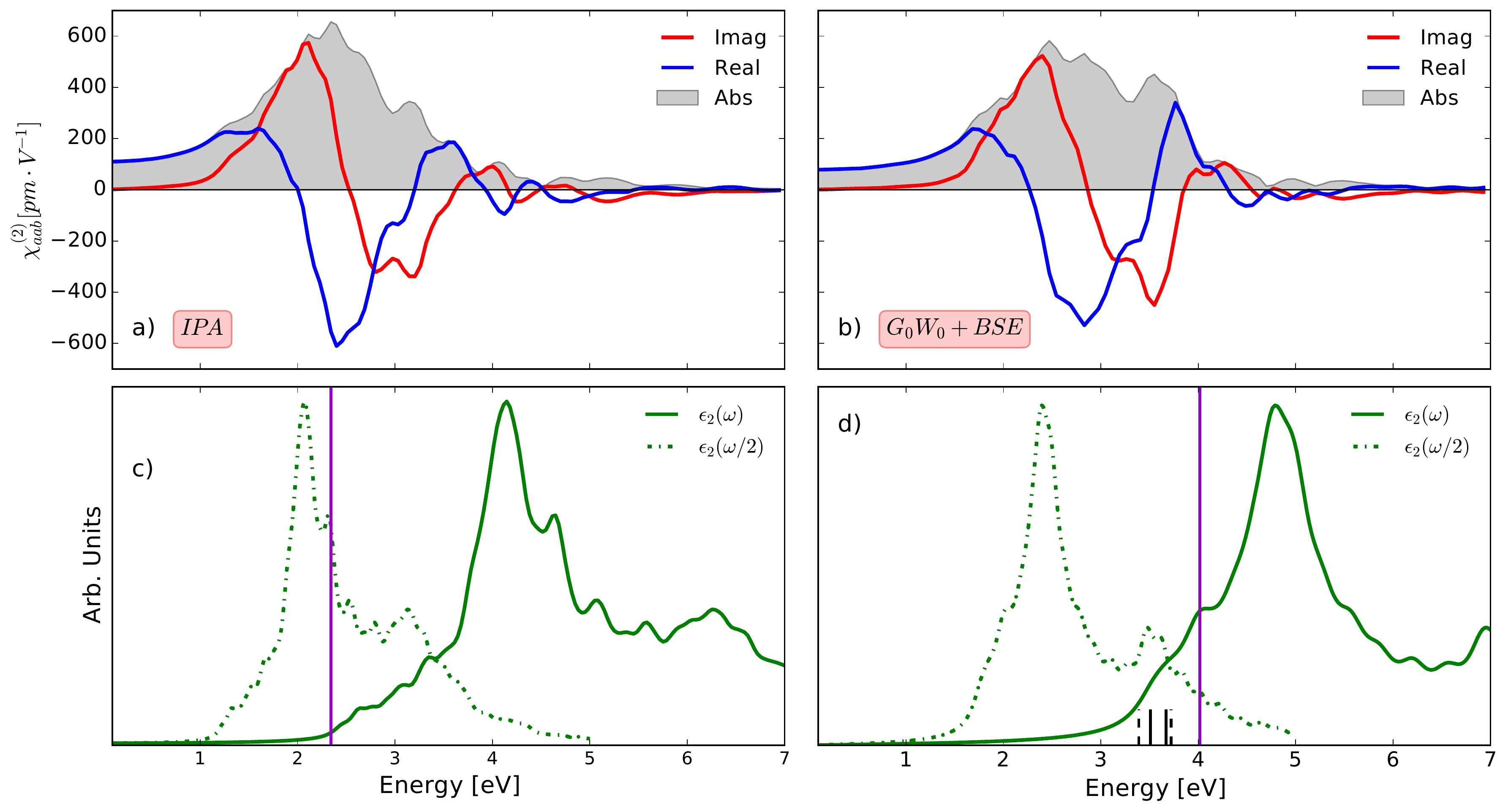}
     \caption{\footnotesize{[Color online] SHG [a) and b)] and optical absorption [c) and d)] of GaSe monolayer for in-plane light polarization calculated at the independent-particle [a) and c)] and at the G$_0$W$_0$ + Bethe-Salpeter Equation level of approximation [b) and d)]. 
     In a) and b) we plot the real part (blue line), imaginary part (red line)
     and the absolute value (overshadow area) of the SHG, $\chi^{(2)}(\omega)$. 
     In c) and d) we plot the imaginary part of the dielectric function calculated at $\omega$ (continuous line) and at $\omega/2$ (dashed line). In d) we also report the position of the two lowest bright (continuous black short vertical line) and two lowest dark (dashed black short vertical line)  exciton energies for in-plane light polarization. The DFT and G$_0$W$_0$ gap are indicated by vertical blue-violet lines.\label{eps}}}
\end{figure*}

\begin{figure*}[ht]
\centering
\includegraphics[width=0.99\textwidth]{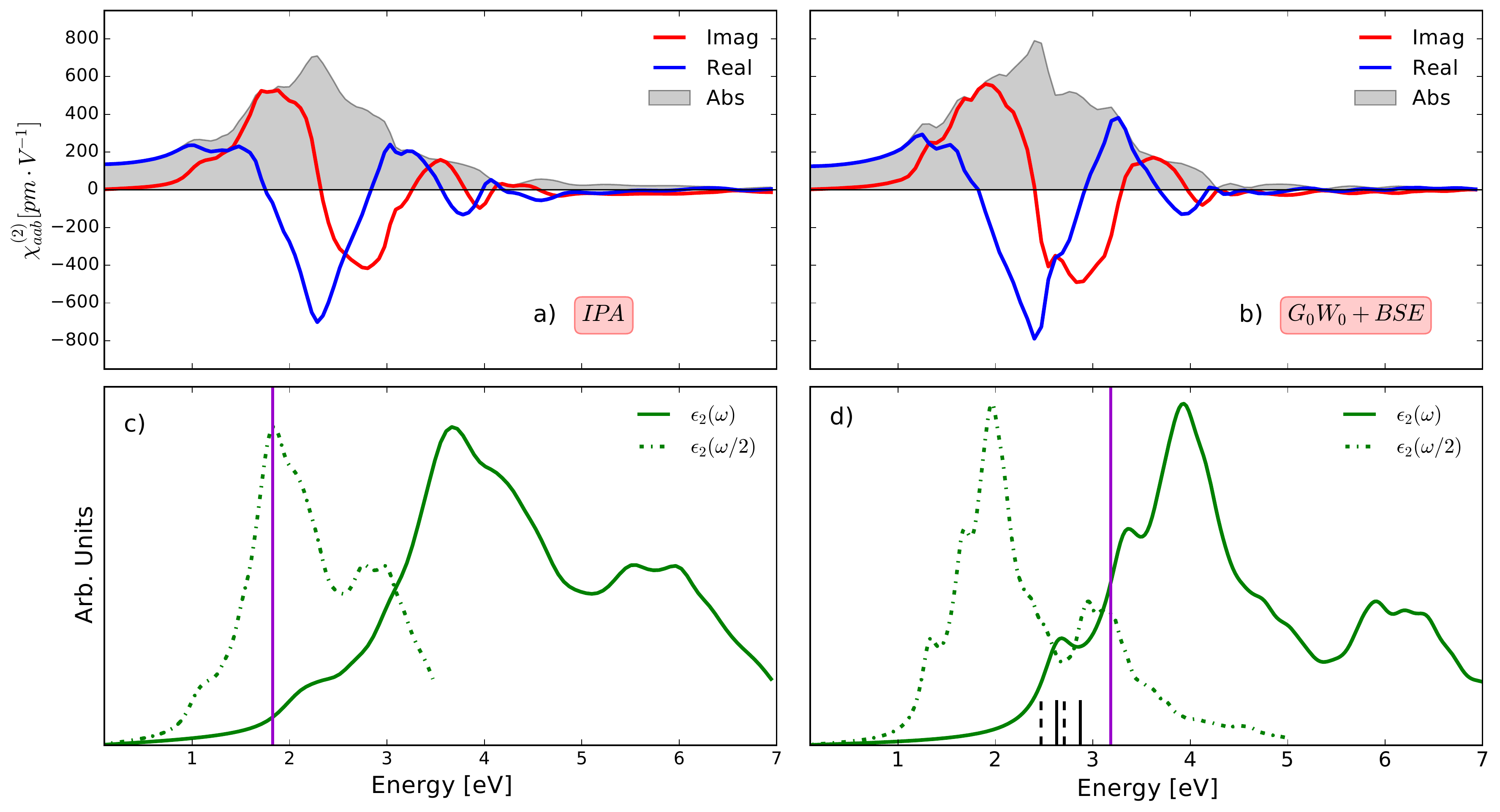}
     \caption{\footnotesize{[Color online] SHG [a) and b)] and optical absorption [c) and d)] of InSe monolayer for in-plane light polarization calculated at the independent-particle [a) and c)] and at the G$_0$W$_0$ + Bethe-Salpeter Equation level of approximation [b) and d)]. 
     In a) and b) we plot the real part (blue line), imaginary part (red line)
     and the absolute value (overshadow area) of the SHG, $\chi^{(2)}(\omega)$. 
     In c) and d) we plot the imaginary part of the dielectric function calculated at $\omega$ (continuous line) and at $\omega/2$ (dashed line). In d) we also report the position of the two lowest bright (continuous black vertical line) and two lowest dark (dashed black vertical line)  exciton energies for in-plane light polarization. The DFT and G$_0$W$_0$ gap are indicated by vertical blue-violet lines\label{epsInSe}}}
\end{figure*}

\begin{table}[h]
\bigskip
    \begin{tabular}{l|c|c|c|c}
        \hline
        &  $\chi^{(2)}$(IPA) & $\chi^{(2)}$ (RPA) & $\chi^{(2)}$ ($G_0W_0$)  & $\chi^{(2)}$ ($G_0W_0$+BSE) \\
        \hline
        GaSe &       110                  &   71.0                   &   54.0          & 81.2       \\
        \hline
        InSe &         135                &    86                  &   68.0 & 122
\end{tabular}
\caption{Static limit of the SHG, $\chi^{(2)}(\omega=0)$, obtained from real-time simulations at different level of approximation: independent-particles approximation (IPA) and the random-phase approximation (RPA) on top of the DFT electronic-structure, the IPA ($G_0W_0$) and the real-time BSE (G$_0W_0$+BSE) on top of the G$_0$W$_0$ electronic-structure. Units in pm/V.}     
\label{tabI}
\end{table}
In Fig.~\ref{eps} and Fig.~\ref{epsInSe} (bottom panels), we report the imaginary part of the dielectric function of GaSe and InSe for in-plane polarized light.
The first excitons, obtained by full diagonalization of the Bethe-Sapeter equations\cite{Strinati1988,lucia} are reported as well. As one can see for both GaSe and InSe we are in presence of bound excitons (i.e. at energies below the fundamental band-gap).
The lowest exciton at 3.39~eV in GaSe is dark for light polarized in the plane due to the symmetry of the bands close to the $\Gamma$ point (while it has a small but non-zero dipole for light polarization in the $z$ direction). The first bright exciton is found at 3.51~eV. Our results are in good agreement with those of Antonius et al.\cite{antonius2018orbital}
For the InSe monolayer (bottom panels of Fig.~\ref{epsInSe}) we obtained similar results. The lowest excitation at $2.47$~eV is dark for in-plane polarized light and the first bright exciton is at $2.63$~eV. This is in agreement with luminescence measurements that found that the lowest-energy transition is not optically active in InSe monolayer.\cite{bandurin2017high}  Therefore, the lowest excitons do not contribute to the second-order susceptibility for in-plane polarized light.


%
%
\subsection{Second Harmonic Generation}\label{sc:res2}
In Table~\ref{tabI}, we report the static $\chi^{(2)}(0)$ at different level of theory: IPA, RPA, independent-particle on top of the G$_0$W$_0$ band structure (G$_0$W$_0$), and finally the full G$_0$W$_0$+BSE. At the independent-particle level we found a large static second order susceptibility for both systems. The susceptibility is sensitive to the electronic structure and in particular to the band gap: a larger band gap corresponds to a smaller $\chi^{(2)}(0)$ as one can see going from DFT to the G$_0$W$_0$. Also, the band structure of InSe is similar to that of GaSe and the gap at DFT level is 1.82~eV against 2.34~eV of GaSe. As a consequence InSe has a larger $\chi^{(2)}(0)$.

The GaSe result at G$_0$W$_0$ level is compatible with the result of Hu et al. that found a $\chi^{(2)} (0) = 69.4$~pm/V starting from a DFT band structure obtained by means of the Heyd-Scuseria-Ernzerhof hybrid functional that has a gap of 3.34~eV, in-between the PBE 2.34~eV and the G$_0$W$_0$ 4.03~eV band gap values. For the InSe monolayer, Lin et al.\cite{lin2018indium} report a value of $91.3$~pm/V, compatible with our result, considering that they assumed a smaller effective thickness. Lin et al. also investigated other possible structures of the InSe monolayer with a stronger SHG but we are not aware of experimental realization of these polymorphs
and therefore we limit our analysis to the hexagonal monolayer. 

Within the RPA, local field effects---that originate from charge fluctuations induced by the system inhomogeneity---attenuate the second order susceptibility in both GaSe and InSe monolayers. This reduction has been observed as well for other low-dimensional systems and in bulk.\cite{Attaccalite2013,Gruning2014} 
Within the G$_0$W$_0$+BSE, excitonic effects tend to increase the $\chi^{(2)}(0)$ and counteract the reduction from both local-field effects and quasiparticle corrections, in agreement with what observed for bulk semiconductors.~\cite{PhysRevB.65.035205} The SHG intensity at $G_0W_0$+BSE level is for both materials closer to the IPA one.

Figures~\ref{eps} and \ref{epsInSe} [(a) and (b) panels] show  the imaginary, real part and absolute value of the SHG for GaSe and InSe monolayers within the IPA (a) and $G_0W_0$+BSE (b). They are compared with the imaginary part of the dielectric function at $\omega$ and $\omega/2$ (bottom panels) at the same level of theory. From the comparison, the main spectral features in the SHG are easily recognizable as two-photon resonances at the van Hove (IPA) and exciton energies ($G_0W_0$+BSE). We focus here on the features below the onset of the optical absorption, that is in the transparency region (imaginary part of the dielectric function $\varepsilon_2$), which is the spectral region of technological interest.
For GaSe, comparing the IPA with the $G_0W_0$+BSE results, we first notice that the transparency region is extended by 1 eV and the SHG intensity (absolute value)  redistributed to higher energies. The lowest energy feature in the SHG intensity at the IPA level is a broad shoulder at about 1.3~eV. In the $G_0W_0$+BSE instead, the lowest feature in the SHG intensity is a  peak at about 1.75~eV which corresponds to a two-photon resonance with the lowest optical active excitons. Overall, the SHG absolute value is lower at the $G_0W_0$+BSE than at the IPA level. The reduction is mainly due to the quasiparticle shift (a larger band gap corresponds to a lower SHG as discussed above) which is only partially compensated by the excitonic effects.
Differently from the GaSe, in the InSe the excitonic effects fully compensate for the reduction of the SHG due to the quasiparticle shift and the local field effects.
The transparency region is extended by 0.4~eV and again the SHG intensity is redistributed at higher energies and enhanced at excitonic resonances. In this case, because of the smaller quasiparticle corrections---the gap opens of 1.35~eV---the SHG at the $G_0W_0$+BSE level is slightly stronger than at the IPA level at resonance. In particular, the lowest energy feature is a broad shoulder at 1~eV within the IPA and a stronger peak at 1.32~eV corresponding to the two photons resonance of the lowest optical active exciton. 

\section{Discussion and conclusions}\label{sc:conc}
The first thing to observe is that the inclusion of many-body effects changes the SHG spectrum: both at the level of the electronic structure (quasiparticle corrections within the $G_0W_0$) and at the level of the response function (inclusion of excitonic effects). Then, many-body effects may be important when considering technological applications (e.g. accurately determining the transparency region and excitonic resonances) and are essential when investigating the physics of excitons of these 2D systems.

On the other hand, the theoretically predicted intensities---both with and without the inclusion of many-body effects---differ substantially from the experimental values. For GaSe, Ref.~\onlinecite{zhou2015strong} reports 0.7~nm/V at 0.77~eV, 1.7~nm/V at 0.92~eV and 2.4~nm/V at 1.02~eV whereas we find 0.096, 0.105, 0.111 nm/V respectively, that is a factor 7--20 smaller. For InSe, Ref.~\onlinecite{zhou2018inse} reports 6.39 nm/V at 1.55 eV whereas we find 0.42 nm/V---that is a factor 15 smaller.

Previously, large differences in the experimental estimate of the SHG coefficients---over a range of three orders of magnitude~\cite{PhysRevB.87.201401,PhysRevB.87.161403,Yin2014}---were observed for the MoS$_2$ monolayer. Eventually, Clark and coworkers~\cite{PhysRevB.90.121409} explained that such differences depend on the model assumed to extract the value of the SHG from the experimental data. Indeed, experimentally the SHG is obtained from the measured second-harmonic intensity. Extracting the SHG coefficient is not straightforward 
and implies to assume a model for the system: Clark and coworkers~\cite{PhysRevB.90.121409} argued that 2D materials should be modelled as a `sheet', as in Refs.~\onlinecite{PhysRevB.87.201401,Yin2014}, rather than as `bulk', as in Ref.~\onlinecite{PhysRevB.87.161403}. Further, they provide an expression for the difference between $\chi_\text{sheet}^\text{(2)}$ and $\chi_\text{bulk}^\text{(2)}$, i.e. the SHG coefficient estimated either using the `sheet' or the `bulk' model:
\begin{equation}\label{eq:shgratio}
    \frac{\chi_\text{bulk}^\text{(2)}}{\chi_\text{sheet}^\text{(2)}}=32\pi\,F \frac{n_\text{2D}(\omega)\sqrt{n_\text{2D}(2\omega)}}{(n_\text{sub}+1)^3},
\end{equation}
where $n_\text{2D}$ and $n_\text{sub}$ are the refractive index for the 2D material and of the substrate. $F$ is an ``overall scaling factor'' that depends on the transmittance and reflectance at the corresponding interfaces air-GaSe (InSe) and GaSe(InSe)-substrate as explained in the Supp. Mat. of Ref.~\onlinecite{PhysRevB.90.121409}.
For MoS$_2$, the ratio in Eq.~\eqref{eq:shgratio} can be as large as 900~\cite{PhysRevB.90.121409} and thus accounts for the differences in the experimental estimates.~\cite{PhysRevB.87.201401,PhysRevB.87.161403} It also indicates that the `bulk' model leads to too large values for the SHG coefficient. This conclusion is also comforted by accurate theoretical predictions~\cite{Gruning2014,Trolle} that reported values close to the experimental estimates that assume the `sheet' model.

Coming back to the monochalcogenides, all the experimental values reported for the GaSe and InSe assumed a bulk model. To investigate whether this is the reason for the discrepancy with the theoretical estimates we use Eq.~\ref{eq:shgratio} with:
$n_\text{sub}\approx 1.45$ (fused silica~\cite{Malitson:65}); $n_\text{GaSe}(\omega)=n_\text{GaSe}(2\omega) = 2.8$,~\cite{zhou2015strong} and $n_\text{InSe}(\omega)$ =  2.86 and $n_\text{InSe}(2\omega)$ =  3.35,~\cite{zhou2018inse}. We set $F=1$, rather than the value of approximately 20 indicated in Ref.~\onlinecite{PhysRevB.90.121409}, since the experimental estimates already account for various gain/loss factors.
We thus obtain ratios as high as $\approx 32$ for GaSe and $\approx 36$ for InSe. These large ratios indicate that, the experimental estimate would be significantly smaller and in substantial agreement with the theoretical predictions if using the `sheet' model.  Residual differences between the experimental and theoretical value could be due to the effects of the substrate and of the surface contributions to the SHG\cite{deckoff2016observing}, which are neglected in the theoretical calculations. Another factor, on the side of the theoretical calculations, could be the chosen dephasing parameter as discussed in Sec.~\ref{sc:theo}. The so-corrected experimental estimate for GaSe is as well closer to the SHG coefficients for the bi- and trilayer,\cite{jie2015layer,PhysRevB.94.125302} which have been obtained by modeling the few-layers as a bulk medium and accounting for interference in the multilayer system.

In summary, using accurate first-principles real-time simulations based on many-body theory, we confirm strong SHG coefficients. 
In particular, for InSe we obtain a SHG coefficient of $\approx 0.5 nm/V$ at 1.67~eV---close to maximum gain and laser efficiency of Ti:Sapphire. In substantial agreement with previous theoretical works, we found that the SHG coefficients are not as strong as claimed in the experimental works of Refs.~\onlinecite{zhou2015strong,zhou2018inse}, differing by an order of magnitude. We argue that the  experimental values are overestimated due to the assumption of a `bulk' rather than a more appropriate `sheet' model to extract the SHG from the measurements and they need to be reviewed. We have shown indeed that a substantial agreement between theoretical and experimental values is recovered when a `sheet' model is used to extract the SHG from the experimental data.

\emph{Note added at resubmission.} Recently, we became aware of a work, Ref.~\onlinecite{doi:10.1063/1.5052417},  which reports experimental measurements of SHG in InSe in good agreement with our theoretical predictions and indeed uses the `sheet' model to extract the SHG coefficient.\\

\section*{acknowledgments}
The research leading to these results has received funding from the European Union Seventh Framework Program under grant agreement no.~696656 GrapheneCore1 and no.~785219 Graphene Core2. M.P. ackowledges ``Tor Vergata'' University for financial support through the Mission Sustainability project 2DUTOPI. E.C.\ acknowledges support by the Programma per Giovani Ricercatori - 2014 ``Rita Levi Montalcini''. M.P., E.C and C.A. acknowledge PRACE for computational resources on Marconi at CINECA. C.A. acknowledges A. Zappelli for the management of the computer cluster \emph{Rosa}. This publication is based upon work from COST Action TUMIEE CA17126, supported by COST (European Cooperation in Science and Technology).

%
\bibliography{gase}

\end{document}